\documentclass[preprint,pre,aps,superscriptaddress,a4paper]{revtex4}

\usepackage[utf8x]{inputenc}
\usepackage{graphicx, subfigure}
\usepackage{epstopdf}
\usepackage{amsmath}
\usepackage{amsfonts}
\newcommand{\ro}{\hat{\rho}}

\begin{document}

\title{Bistability in a self-assembling system confined by elastic walls. Exact results in a one-dimensional lattice model.}
\author{J. P\c ekalski}
\affiliation{Institute of Physical Chemistry,
 Polish Academy of Sciences, 01-224 Warszawa, Poland}
\author{ A. Ciach}
\affiliation{Institute of Physical Chemistry,
 Polish Academy of Sciences, 01-224 Warszawa, Poland}
 \author{N. G. Almarza}
\affiliation{Instituto de Qu{\'\i}mica F{\'\i}sica Rocasolano, CSIC, Serrano 119, E-28006 Madrid, Spain }
\date{\today}
\begin{abstract}
The impact of confinement on self-assembly of  particles interacting with short-range attraction  and long-range repulsion (SALR) potential
is studied for thermodynamic states corresponding to local ordering of clusters or layers in the bulk.
Exact and asymptotic expressions for the local density and for the effective potential between 
the confining surfaces are obtained for a one-dimensional lattice model introduced in
[J. P\c ekalski et. al. \textit{J. Chem. Phys.} {\bf 140}, 144903 (2013)].
%that concerns an open system (permeable boundaries). 
The simple asymptotic formulas are shown to be in good quantitative agreement with
exact results for slits containing at least 5 layers. 
We observe that the incommensurability of the system size and the average distance between the clusters or layers in the bulk 
leads to structural deformations that are different for different values of the chemical potential $\mu$. 
The change of the type of defects 
is reflected in the dependence of density on $\mu$ that has a shape characteristic for phase transitions. 
Our results may help to avoid misinterpretation of the
change of the type of defects as a phase transition  in simulations of inhomogeneous systems. 
 Finally, we show that a system confined by soft elastic walls may exhibit  
 bistability such that two system sizes
 %, corresponding to the numbers of layers 
 that differ approximately by the average distance between the clusters or layers 
 %in the bulk 
 are almost equally probable. 
 This may happen when
 the equilibrium separation between the soft boundaries of an empty slit corresponds
 to the largest stress in the confined self-assembling system.
\end{abstract}
\maketitle
\section{Introduction}
Self-assembly  is a fundamental process in  living matter, it is even thought to be the key to understand
the origin of life as it leads to organization of intracell compartments \cite{souza:11:0}. 
Self-assembly is responsible for spontaneous formation of lipids into a bilayer. Moreover, it      
takes part in intercell communication when proteins adsorbed on a cell membrane spontaneously aggregate
and form clusters which can perform functions unavailable to a single protein molecule \cite{lang:10:0,Saric:13:0,Stradner:04:0}.
On the other hand, the self-assembly of nanoparticles is of high interest in industry. In nanotechnology self-assembly 
is a base for techniques aimed at device miniaturization and material production with novel electronic, 
mechanical and optical properties \cite{Hu:14:0,Kitching:13:0}. 

% \cite{Lang10} - review - Membrane Protein Clusters at Nanoscale Resolution: More Than Pretty Pictures (ex)
% \cite{Saric13} - review self-assembly on membranes (theory)
% \cite{Stradner04} - Equilibrium cluster formation in concentrated protein solutions and colloids
% \cite{Hu14} - review direct self-assmebly of block copolymers in nanotechnology (ex)
% \cite{Palmer08} - Molecular Self-Assembly into One-Dimensional Nanostructures (Ex)
% \cite{Kitching13} - review Self-assembly of metallic nanoparticles into one dimensional arrays (ex techniques)
% \cite{Gao} Imaging enzyme-triggered self-assembly of small molecules inside live cells 2012 nature http://www.ncbi.nlm.nih.gov/pmc/articles/PMC3521559/
% \cite{Souza} Spontaneous Crowding of Ribosomes and Proteins inside Vesicles: A Possible Mechanism for the Origin of Cell Metabolism, DOI: 10.1002/cbic.201100306
% \cite{Godfrin14} Generalized phase behavior of cluster formation in colloidal dispersions with competing interactions, SM 2014
Heterogeneity on a mesoscopic length scale is often an effect of competing tendencies in the pair interaction potential.
In the case of nanoparticles or globular proteins  there is a competition between solvent-induced attraction,
and repulsion that is typically (but not exclusively) of electrostatic origin \cite{Stradner:04:0,campbell:05:0,shukla:08:0}.
The effective isotropic short-range attraction (SA) between nanoparticles, ions or organic molecules favours their 
aggregation, while the presence of the long-range isotropic repulsion (LR) effects in the separation of the aggregates. 
The competition between these opposite tendencies results in thermodynamic stabilization of  spatially 
inhomogeneous patterns made of globular or elongated clusters, or layers (stripes in a case of a surface). 
The clusters or layers are periodically distributed and form regular patterns in ordered 
phases~\cite{archer:08:0,imperio:06:0,ciach:08:1,ciach:10:1}.  In the disordered phase
the particles also self-assemble into clusters or layers
 for some range of  temperature and concentration~\cite{toledano:09:0,kowalczyk:11:0,almarza:14:0}. 
However, in the disordered phase these objects are ordered only locally, and this short-range order is
reflected in the exponentially damped oscillatory 
behavior of the correlation function on the mesoscopic length scale~\cite{ciach:08:1,pekalski:13:0,pekalski:14:0,archer:08:0}.

In intracell compartments, in pores of a porous material or on geometrically patterned surfaces, the soft or rigid 
boundaries of the system can have an ordering or disordering effect on the confined clusters or layers.
The key factor is the commensurability of the typical distance
between the objects in the bulk, and the size of the compartment. Despite
the fact that the confinement plays a very important role in biological systems, in pores of  porous materials, and on 
patterned surfaces,
the effects of confinement on the self-assembling systems have been much less studied than the bulk properties. 
In the case of the  SALR interaction potential 
the impact of a slit-type  confinement on   thermodynamically stable patterns on a surface (2d system) was studied 
by Monte Carlo simulations \cite{imperio:07:0}
and by density functional theory \cite{archer:08:0}. In Ref.~\cite{imperio:07:0}
the authors found that unlike in the bulk system, in presence of neutral walls a switch from the cluster to the lamellar morphology 
with increasing temperature is possible. Moreover, the orientation of the lamella depends on the distance 
between the walls and the particle-wall interaction parameters. In  Ref.\cite{archer:08:0} 
the author focused on determining the sequences of stable structures for increasing distance between the walls 
at a given temperature and for fixed density.
He confirmed that the change of the distance can lead to the change of the stable-phase morphology, especially if
the period of the structure stable in the bulk and the width of the slit are incommensurate.

%The role of balance between entropy and energy in determination of stripes orientation is left as an open question.

In this work we focus on these effects of confinement on the SALR systems that  have not been investigated yet,
although in our opinion play 
a very important role. 
We limit ourselves to the disordered phase, and consider only
permeable confining walls, i.e. the system can interchange particles with a reservoir (Grand canonical ensemble). 
Our first question is how the 
structural defects in the case of incommensurability between the system size and the period in the bulk phase 
depend on thermodynamic state
and on the interaction with the surfaces. The second question concerns the fluid-induced effective 
interactions between the confining surfaces
for different values of the chemical potential. The  periodic order on the mesoscopic length scale can induce
a periodic effective interaction between confining surfaces  on the same length scale. This is analogous to
the periodic solvation force on the atomic length scale in simple fluids \cite{israelachvili:76:0,israel:11:0}. 
In contrast to the amphiphilic systems, 
where the effective interaction  was intensively investigated both experimentally
\cite{kekicheff:89:0,kekicheff:95:1} and 
theoretically~\cite{matsen:96:1,matsen:97:0,tasinkevych:99:1,tasinkevych:99:0,babin:01:0,babin:03:0},
 in the case of the SALR potential it has not been studied yet.

In  biological systems or in pores of a soft porous material 
the compartments are surrounded by lipid bilayers or by elastic material. The 
separation between the confining surfaces can be varied, and this change is associated with  elastic energy. 
 Mechanical equilibrium between the solvation force resulting from the stress in the confined self-assembling system,
 and the elastic force resulting from  deformation of the confining elastic material (e.g. lipid bilayer)
determines the equilibrium shape of the system boundary. 
Changes of a thermodynamic state, leading to the change of the solvation force, might lead
to shape and/or size transformations.
In this work we address a question of equilibrium wall separation in a pore of an elastic material
containing particles interacting with the SALR potential.
It would be interesting to know the effect of the presence of particles on the equilibrium separation between the elastic walls, specially when the equilibrium thickness of the pore in the absence of particules and the period of the bulk phase are incomensurate.

In order to address the aforementioned questions on a general level we consider a generic lattice model of a SALR system
that can be solved exactly in one dimension (1d). Our results can give some insight for the properties 
of two and three dimensional systems confined
in slits. Moreover, 
a pseudo-1d system confined by elastic boundaries is formed, for example, by a long protrusion in a vesicle 
filled with charged nanoparticles. In Ref.~\cite{gozdz:05:0} it was shown that such a protrusion responds elastically 
to an external stress in direction parallel to its axis.
The bulk properties of the model were thoroughly studied
in Ref.~\cite{pekalski:13:0}.
In 1d the ordered phases appear only at $T=0$. The short-range order and the pretransitional effects, however, can be
studied based on exact solutions. We found that in the disordered phase 
the repulsion between the clusters or layers leads to a dependence of the average density $\rho$ on the chemical potential $\mu$ or pressure $p$ 
that is significantly different to that of simple fluids \cite{pekalski:13:0}.
A characteristic plateau in $\rho(\mu)$ and $\rho(p)$ appears
when the density is equal to the density of the periodic structure that is favoured energetically.
This plateau signals that a significant increase of pressure is necessary
to overcome the repulsion between the clusters and to compress
the system to  a dense structure. In addition,  for the range of $\mu$ corresponding to the plateau in $\rho(\mu)$
the correlation length
is 3 or 4 orders of magnitude larger than the size of the particles.
The shape of the  $\rho(p)$ curve and the large correlation length suggest that 
the solvation force can be quite strong in large pores even in the case of the short-range periodic order. 
We shall verify this expectation by exact results.

In sec.2 we briefly describe the model introduced in Ref. \cite{pekalski:13:0} and the transfer matrix method. 
The details of the derivations are described in Appendices. In sec.3 we present 
asymptotic expressions for the local density
and for the effective potential between the confining surfaces. We determine the range of validity of these formulas
by comparison with exact results. In sec. 4 we discuss the dependence of the distribution of the particles inside the pore
on the chemical potential when 
the width of the slit and the period of the bulk structure are incommensurate. We also compare the shape of $\rho(\mu)$ 
for various slits with the result obtained in Ref.\cite{pekalski:13:0} for the bulk.
In addition,
%to the slit with various kinds of surfaces, 
 we consider periodic boundary conditions in the case of incommensurability, in order to  help to interpret simulation results. 
 The effective interaction for different thermodynamic states and the equilibrium width of a system with  elastic boundaries 
 are determined in sec.5. The summary and conclusions are 
 presented in sec.6. \newline

\section{The model and the method of exact solutions}
\begin{figure}[th]
\centering
\includegraphics[scale=1]{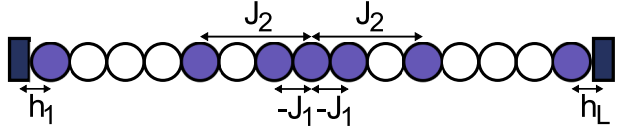}
\caption{Scheme of the model for a system of size $L=15$. The lattice constant $a$ is equal to the particle diameter $\sigma$. 
The particles attract or repel each other with the energy $-J_1$ or  $J_2$ when they are the nearest or the third neighbors 
respectively. If a particle occupies the first or the last site of the lattice then it interacts with the
confining wall with the energy $h_1$ or $h_L$ respectively.}
\label{model}
\end{figure}

We consider a lattice model for systems with particles interacting with a short-range attraction and long-range
repulsion (SALR) potential. We assume that the particles 
occupy lattice sites on a 1 dimensional lattice, and the lattice constant $a$ is comparable with the particle diameter $\sigma$.
The particles attract or repel each other when they are the nearest or
the third neighbors respectively. The model with the same interaction potential and with periodic boundary
conditions (PBC) was solved exactly in Ref.\cite{pekalski:13:0} for the system sizes $L=6N$, where $N$ is integer.
For such system sizes the energetically favourable structure of 3 occupied sites separated by 3 empty sites 
is possible, and  the properties of the bulk system can be reproduced. For $L\ne 6N$  the 
 incommensurability between the system size $L$ and the period of the ordered structure may influence the results.
Here we focus on the  impact of this incommensurability in the case of the PBC and in the case of a slit  type of confinement,
i.e. with rigid boundary conditions (RBC). In the case of the PBC we are interested in how the 
incommensurability influences the dependence of the density $\rho$ on the chemical potential $\mu$ for different system sizes.
Our exact results may 
help to interpret the results of simulations that are performed for finite systems when the incommensurability might be a serious issue.
%in 2 and 3 dimensional systems.  
In the case of the RBC we assume that the confining walls can interact with the 
particles located 
at the first and the last site of the system  (see Fig. \ref{model}). The confining walls represent  real physical confinement
e.g. in a porous material or in a thin film on a solid substrate.

We assume that the lattice consists of $L$ sites labeled from $1$ to $L$. In order to describe whether the lattice site 
$x$ is occupied or not we  introduce an occupation operator $\hat\rho(x)$ which is equal to 1 or 0 respectively. 
Hence, the configuration of the system (the microstate) is given by $\{\hat\rho(x)\}\equiv(\hat\rho(1),...,\hat\rho(L))$. 
The probability of the microstate  $\{\hat\rho(x)\}$ is
\begin{equation}
\mathbb{P}[\{\hat\rho(x)\}]=\frac{e^{-\beta H[\{\hat\rho(x)\}]}}{\Xi},
\end{equation}  
where $\Xi$ is the Grand Partition function, $\beta=(k_BT)^{-1}$, $k_B$ is the Boltzmann constant
and $H$ is the thermodynamic Hamiltonian which
contains the energy and the chemical potential term
\begin{equation}
\label{H}
  H [\{\hat\rho\}] = \frac{1}{2} \sum_{x=1}^L\sum_{x'=1}^L \hat{\rho}( x) V ( x-x')\hat{\rho}(x') + h_1 \hat{\rho}(1) + h_L \hat{\rho}(L)
- \mu \sum_{x=1}^L \hat{\rho}( x),
  \end{equation}
 where  the particle-particle interaction potential is

\begin{equation}
\label{V}
V(\Delta{ x}) = \left\{ \begin{array}{ll}
-J_1 & \textrm{for $|\Delta{ x}| = 1$},\\
+J_2 & \textrm{for $|\Delta{ x}| = 3$},\\
0 & \textrm{otherwise.}
\end{array} \right.
\end{equation}

We choose $J_1$ as the energy unit and introduce dimensionless variables for any quantity $X$ with 
dimension of energy as $X^*=X/J_1$, in particular
\begin{eqnarray}
\label{dimensionless}
T^*&=&k_BT/J_1,\quad J^*=J_2/J_1,\quad h_1^*=h_1/J_1,\quad \\ h_L^*&=&h_L/J_1,\quad \mu^*=\mu/J_1.
 \end{eqnarray}
 
 We solve the model exactly by the transfer matrix method, as in Ref.~\cite{pekalski:13:0}. 
 Because the range of the particle-particle interactions is $3$, 
 the transfer matrix operates between the microstates
 in boxes  that are located next to each other, and each box consists of 3 sites. There are 8 microstates in each box, therefore
 the dimension of the transfer matrix is 8. The system can be divided into such boxes when
 $L=3N$. In general, the expression for the  Grand Partition function for $L=3N+j$ depends on both, $N$ 
 and the reminder of division of L by 3, $j = L \, mod \,\, 3 =0,1,2$. 
 We describe the method in more detail, and give the exact expression for $\Xi$ in Appendix A.
 
 When the correlation length between the particles is comparable with the distance between the confining walls, then the
distribution of the particles is influenced by both walls. This leads to the excess of the grand potential depending on the
distance between the walls \cite{evans:90:0}, 
 %The thermodynamic effects of confinement are described by the excess grand thermodynamic potential
 \begin{equation}
 \label{Omex}
  \Omega_{ex} \equiv \Omega - \Omega_{bulk} = \gamma_1 + \gamma_L + \Psi(L)
 \end{equation}
where $\Omega=- k_BT \ln \Xi$ and $\Omega_{bulk}=- k_BT \ln \Xi_{bulk}$ are the grand potential
 in the slit and in the bulk of the same size $L$ respectively, $\gamma_1 $ and $\gamma_L$ 
 are the wall-fluid surface tensions and $\Psi(L)$
 corresponds to the effective interaction between the confining walls~\cite{evans:90:0}. 
 The  effective force between the surfaces is $-\nabla\Psi(L)$.
 The exact expressions for $\gamma_1,\gamma_L$ and $\Psi(L)$ are given in Appendix C.

 The expression for the local average density at the site $x=3n+l$ in the system  of size $L = 3N+j$  with $j=0,1,2$ 
 depends on both,  $n$ and $l = 1,2,3$,
 and in addition  on  $N$ and $j$. The rather complex formulas are given in Appendix B.
 The exact expressions take much simpler asymptotic forms for $N\gg 1$ and $n\simeq N/2$. We present the asymptotic formulas
 for $N\gg 1$ for the density and for $\Psi(L)$,
 and compare them with the exact results in the next section.

 \section{asymptotic expressions for large slits and the range of their validity}
 
In the energetically favourable structure clusters
 composed of 3 particles are separated by 3 empty sites. For this reason the properties of the  system confined in  
 the slit of large width $L=3N+j$ depend on both,  the number of the triples of sites,  $N\approx L/3$, and
 the number of the additional sites, $j=0,1,2$.
 Let us first consider 
 the average local density in the central part of the slit of large width, $N \gg 1$. 
 We divide the system into triples of sites. 
 Each site $x$ is characterized by the number of the triple to which it belongs, $n$, 
 and the position inside the triple, $l$, so that  $x=3n+l$ with $l=1,2,3$.
 The expression for $\langle \hat\rho(3n+l)\rangle$ depends on $n$ and $l$, as well as on  $N$ and $j$.
From the exact formulas given in  Appendix B we obtain 
 the asymptotic expression for $N\to\infty$ and $n\simeq N/2$ (central part of the slit)
\begin{eqnarray}
\langle\hat\rho(3n+l)\rangle &\simeq&  \bar\rho  + A_1 (l) \cos(n \lambda\! +\! \theta_{1}(l)) e^{-3n/\xi}
%\\
%\nonumber
+
A_L(l) \cos((N-n)\lambda\! +\! \theta_{L}(l)) e^{-3(N-n)/\xi}.
\label{avdenapp}
 \end{eqnarray}
 The explicit expressions for $\bar\rho$,
 %$\langle\hat\rho\rangle $, 
 the  amplitudes  $ A_1(l),A_L(l)$ and the phases  $\theta_1(l),\theta_L(l) $ 
 are given in  Appendix B (these quantities depend also on $N$ and $j$). 
 The decay length $\xi$ is given by the same expression as the correlation length in the bulk \cite{pekalski:13:0},
\begin{eqnarray}
\label{xi}
 \xi=3/\ln\Bigg(\frac{\lambda_1}{|\lambda_2|}\Bigg),
\end{eqnarray}
where $\lambda_1$ and $\lambda_2=|\lambda_2|\exp(i\lambda)$ are the largest and the second largest eigenvalues of the transfer matrix.
 The transfer matrix is not Hermitian, and some of the eigenvalues can be complex. The 
 presence of the imaginary part of $\lambda_2$ depends on $J^*$, and on the thermodynamic state. 
The monotonic decay of the density near a single surface occurs  when $\lambda_2$ is real and positive ( $\lambda=0$). 
The exponentially damped periodic structure with the period 6 occurs  when $\lambda_2$ is real and negative,
($\lambda=\pi$). In most cases, however, including $J^*=3$ for the range of $\mu^*$ studied in this work, 
$\lambda_2$ is complex and the period of 
the damped oscillations
is noninteger.

The asymptotic formula for the effective interaction potential for  $N\to\infty$ is
\begin{eqnarray}
\beta\Psi(3N+j) \simeq   A(j)\cos (\lambda N+\phi(j)) e^{-3N/\xi}.
\label{psi_ap}
\end{eqnarray}
 The explicit expressions for  the amplitude  $ A(j)$ and the phase  $\phi(j) $ 
 are given in  Appendix C.

The asymptotic formulas are simply
the exponentially damped periodic functions.
%with the period 3 or 6 for $\lambda_2>0$ or $\lambda_2<0$ respectively(Eqs. (\ref{P1}) and  (\ref{ro1})). In
 %addition,  modulations with a noninteger wavenumber occur  for complex $\lambda_2$ (Eqs.(\ref{psi_ap}) and  (\ref{avdenapp})).
Similar expressions, but without the amplitude modulations, were obtained in mean-field theories of 
confined self-assembling systems \cite{ciach:01:2,ciach:04:0,schmid:93:0,tasinkevych:99:0}.
These rather simple asymptotic forms are strictly valid for $N\gg 1$ and $n\simeq N/2$. 
 We check the validity of the asymptotic expressions
by comparing them with the exact results. The exact and asymptotic formulas are valid for any $J^*$.
To fix attention 
  we focus  in this paper on the case of  strong repulsion, $J^*=3$, in the analysis of mechanical and structural 
  properties of the confined self-assembling system.

\begin{figure}[th]
\centering
\includegraphics[scale=1]{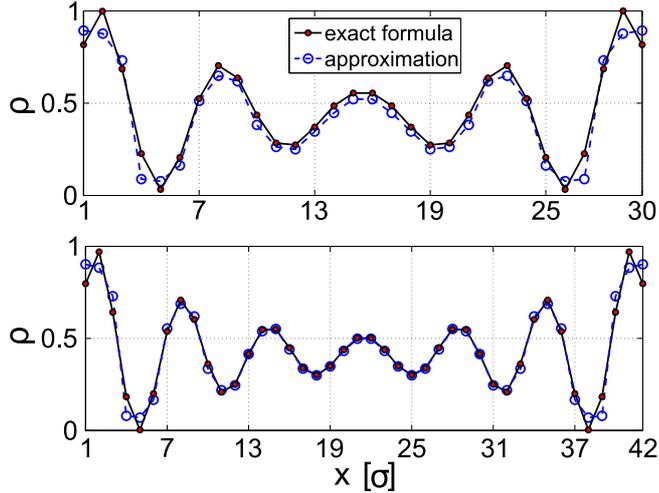}
\caption{Comparison of the exact (\ref{avdenex}) and approximate (\ref{avdenapp}) formulas
for the average density for $J^*=3$, $\mu^*=0, T^*=0.5$ and $h^*_1=h^*_L=-1$.
Upper panel $L = 30$, lower panel $L=42$.} 
\label{ex_approx}
\end{figure}

As shown in Fig.\ref{ex_approx}, the agreement of the asymptotic expression for the local density with 
the exact result is very good already 
for $L=42$, and 
the discrepancy between the exact and asymptotic expressions appear only very close to the surface. For $L=30$ 
the accuracy of the asymptotic expression is less good but it is still satisfactory, 
except from the  clusters adsorbed at the surfaces, where 
some discrepancy can be observed. Thus, the asymptotic formula is sufficiently accurate
not only in the center, but inside the whole slit for slits containing  5
or more clusters. 

\begin{figure}[th]
\includegraphics[scale=1]{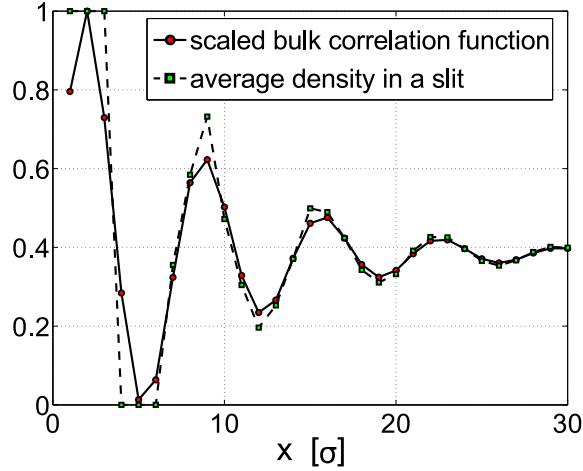}
\caption{Comparison of the  density profile in a slit  (black dashed line) 
for  $J^*=3$, $\mu^* = -0.66$, $T^* = 0.0125$, $h^*_1=h^*_L=-1$  and $L = 96$, and the bulk correlation function obtained in \cite{pekalski:13:0} (red solid line)
for the same thermodynamic state.
The correlation function was  linearly scaled  and shifted by the average density of the bulk system, $\rho = 0.3841$.}
\centering
\label{corr_den}
\end{figure}

In the asymptotic expressions the decay length and the period of oscillations of the local density in the slit and the correlation 
function in the bulk are the same. In Fig.\ref{corr_den} we compare the exact results for the local density  and for the correlation function. 
In order to compare the two functions, we  add the average density of the bulk system to the linearly scaled correlation 
function, and obtain good agreement 
for the distance from the surface $z>10$. We conclude that the correlation function in the bulk describes very well the local structure 
(up to an amplitude that depends on the kind of the wall) except for the first 
cluster adsorbed at the surface.

\begin{figure}[th]
\includegraphics[scale=1]{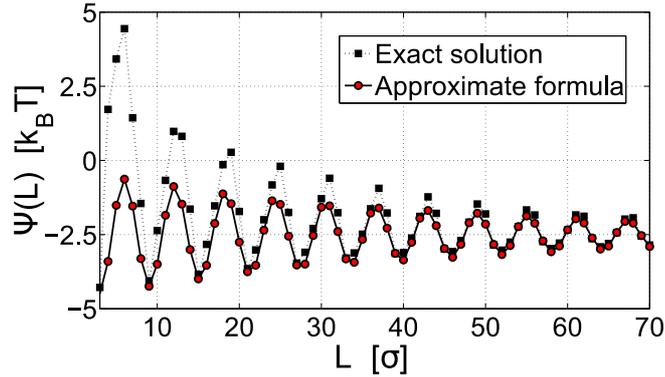}
\caption{The effective wall-wall interaction $\Psi(L)$ for  $J^*=3$, $\mu^*=0$, $T^* = 0.2$ and $h^*_1=h^*_L=-1$. 
Dashed line with squares -
 the exact formula,
Eq. (\ref{Psiex}). Solid line with red circles - the approximate formula, Eq. (\ref{psi_ap}).}
\label{psi_approx}
\end{figure}

In Fig. \ref{psi_approx} the exact and approximate results for the effective interaction between the walls 
%$\Psi$ 
are compared.
As expected, the accuracy of the asymptotic formula improves with increasing system size. 
Close to the minima, i.e. near the equilibrium separations between the surfaces,
the approximate formula works well also for small systems.
On the other hand, for small and incommensurate system sizes the approximate formula highly underestimates 
the interaction potential, therefore for small systems it underestimates the effective force between the walls. 

Notice that for some system sizes there is no clear minimum and $\Psi(L)$ takes almost equal values 
for two consecutive system sizes. However, for the exact and the approximate 
formulas this phenomenon occurs for different system sizes  eg. in Fig. \ref{psi_approx} for $L=39,40$ for
the exact result, and for $L = 27,28$  for the approximate formula.

\section{Effects of incommensurability of the system size and the period of the bulk structure}

In this section we study the effect of the incommensurability of the system size and the period of the bulk structure
on the distribution of the particles and on the dependence of the average density on the chemical potential. 
 Our aim is to verify how $\rho(\mu^*)$
is influenced by the presence of structural defects that must be present in the case of the incommensurability.
We first consider the PBC, and next the RBC.

\subsection{The case of periodic boundary conditions (PBC)}

We  focus on $L=6N+3$, i.e. on
the largest mismatch between $L$ and 6 (the low-$T$ period in the ordered phase).
\begin{figure}[th]
\includegraphics[scale=1]{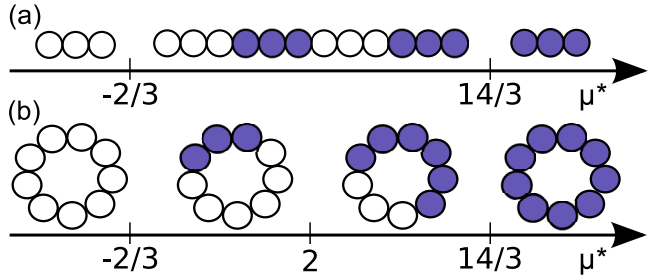}
\caption{Scheme of the ground state ($T^* = 0$) for $J^*=3$ for the bulk system ($L=6N$) (a) and for the system of size $L=9$ (b).
For $\mu^* <-2/3$ or  $\mu^*>14/3$ the stable phase is the vacuum or  the fully occupied lattice. For $-2/3<\mu^*<14/3$ 
the  periodic phase is stable in the bulk. For $L=9$ the stability region  of the periodic phase is split into 
$\mu^*<2$ and $\mu^*>2$ corresponding to 
enlarged void or  cluster respectively. }
\label{gspbc}
\end{figure}
Let us first investigate the ground state (GS), i.e. the case of  $T^*=0$.
For $L=6N+3$ we may expect that in the periodic phase either a separation between some clusters is larger than 3,
or some clusters are larger than 3
 (see Fig.\ref{gspbc}). 
When the separation between the clusters is larger than 3 and we add one particle to a  cluster consisting of at least 3 particles,
then 
the increase of the Hamiltonian is $\Delta H^*=-1+J^*-\mu^*$. For $\mu^*<J^*-1$ or  $\mu^*>J^*-1$ 
we have $\Delta H^*>0$ or $\Delta H^*<0$  respectively. Thus, in the GS corresponding to the minimum of the Hamiltonian $H^*$
the voids in the first case and the clusters in the second case
occupy 3 more sites. At $T^*=0$ the average density jumps by $3/L$ for  $\mu^*=J^*-1$.  
  The GS in the bulk ($L=6N$) and for $L=9$ is shown in Fig.\ref{gspbc} for $J^*=3$.  Note that we have 
  $\Delta H^*=-1+J^*-\mu^*=0$ for $\mu^*=2$ in this case, therefore for $\mu^*=2$ the GS is degenerate, and the
cluster  can consist of either 3,4,5 or 6 particles.

\begin{figure}[th]
\centering
\includegraphics[scale=1]{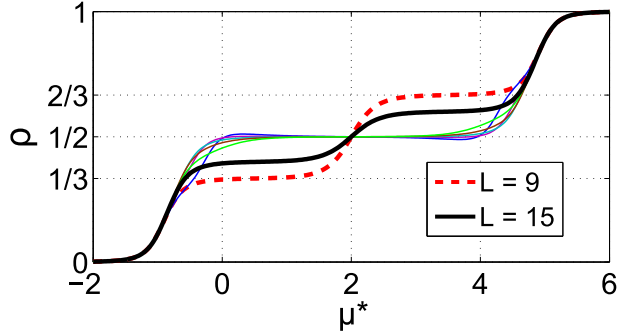}
\caption{The average density $\rho^*$ as a function of the chemical potential $\mu^*$ for  $J^*=3$ and $T^* = 0.3$, 
for a system with PBC and $L = 9$ (red dashed line), $L=10 \ldots 14$ (thin color lines) and  $L = 15$ (black solid line).
 }
\label{PBC_walls}
\end{figure}

In Fig.\ref{PBC_walls} we show $\rho(\mu^*)$ for $T^*=0.3$. Note that the transition between the two types of defects, i.e.
larger voids or larger clusters could be misinterpreted as a transition between different phases, because
when a system undergoes a first-order phase transition, steps in $\rho(\mu^*)$ appear. Thus, the results of simulations 
in the case of systems with spatial inhomogeneities on a mesoscopic length scale should be
interpreted with special care, 
especially when several periodic phases with different periods can appear. However, in the case of structural
defects the height 
of the step decreases as $\sim 1/L$ for increasing $L$, and for certain system sizes the step  disappears (Fig.\ref{PBC_walls}). 
This observation  may help to interpret the simulation results.

\subsection{The case of rigid boundary conditions (RBC)}

\begin{figure}[th]
\includegraphics[scale=1]{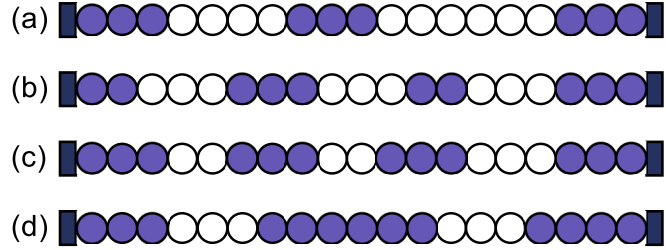}
 \caption{Typical microstates in  the  degenerate GS for a slit of size $L=19$ with attractive walls. 
The range of the chemical potential corresponding to the shown microstates  is (a) $-2/3<\mu^*<0$, (b) $0<\mu^*<J^*-1$,
 (c) $J^*-1<\mu^*<2(J^*-1)$ and (d) $2(J^*-1)<\mu^*<2J^*-4/3$.
}
 \label{L19GS}
\end{figure}

We first focus on attractive surfaces. Let us consider the Hamiltonian for a single cluster composed of $n\le 3$ particles 
adsorbed at the surface, 
$H^*= h^*_1 -(n-1) -n\mu^*$. The adsorption of the cluster is energetically favourable compared to vacuum 
for $\mu^*>(h^*_1 +1-n)/n$.  In order to fix attention, we assume that the interaction with the surfaces is the 
same as the particle-particle attraction, 
$h^*_1=h^*_L=-1$. In this case  a cluster  adsorbed at each  attractive surface is energetically favourable for $\mu^*>-1$.
Thus, for  $\mu^*>-1$  the largest mismatch between the system size and
the structure of the bulk  periodic phase
occurs for $L=6N$ when both  surfaces are attractive.
 For $L\ne 6N+3$ 
the GS of the system  is  degenerate 
in the whole stability region of the periodic phase, because the defects in the periodic structure that are caused by the 
incommensurability of the period and the system size are not localized. Moreover,
%in the slit
the stability region of the periodic phase 
splits into 4 regions, corresponding to different numbers and sizes of the clusters present in the slit.
We choose  $L=19$ and present  typical microscopic states of the GS in Fig.\ref{L19GS}. 
For $\mu^*<0$ there are 3 clusters in the slit. Each of them consists of 3 particles, and the neighboring 
clusters are separated by at
least 3 empty  sites. Apart from this limitation the position of the central cluster can be arbitrary. 
For $0<\mu^*<J^*-1$ there are 2 clusters consisting of 3 particles and 2 clusters consisting of 2 particles in the slit.
The clusters do not repel each other, i.e. there are 3 empty sites between the neighboring clusters. For  $J^*-1<\mu^*<2(J^*-1)$
there are 4 clusters consisting of 3 particles. Finally, for  $\mu^*>2(J^*-1)$ there are 3 clusters 
separated by two voids composed of 3 empty sites, and each cluster consists
of at least 3 particles. 

\begin{figure}[th]
\includegraphics[scale=1]{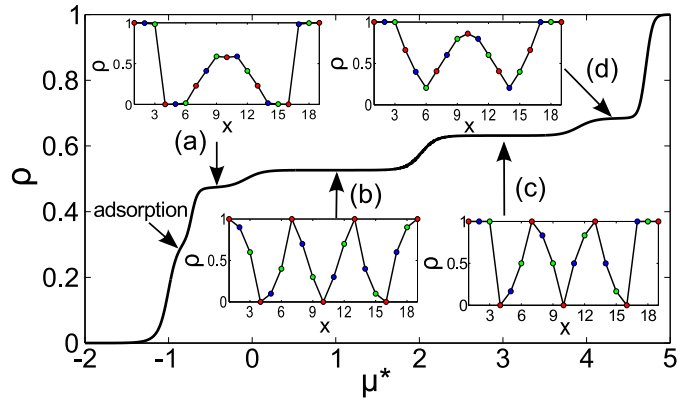}
\caption{Density $\rho$ as a function of the dimensionless chemical potential $\mu^*$  for $J^*=3$ and $T^* = 0.1$ in a slit
of size $L = 19$ 
with  attractive walls. For increasing $\mu^*$ we first observe the 
 adsorption, and next 4 plateaus. The plateaus from (a) to (d) correspond to the average densities shown in the insets.
 The steps between them occur for $\mu^*\approx 0,  2, 4$, i.e. near the GS coexistence between different structures in confinement
  (see Fig.\ref{L19GS}).
 }
\label{L19_overview}
\end{figure}

In Fig.\ref{L19_overview} we present $\rho(\mu^*)$ for the slit with  $L = 19$ at $T^* = 0.1$. 
The average densities corresponding to the plateaus are shown in the insets. 
Note the similarity between the average densities in the GS (a)-(d) (Fig.\ref{L19GS}) and  the insets (a)-(d) in 
Fig.\ref{L19_overview}. 

In the case of  confinement in a slit the steps in $\rho(\mu^*)$ represent a physical 
effect, namely  structural changes such as a jump of a number of the clusters or a change of their size 
as a function of the chemical potential. Such abrupt changes in a slit  induced by small changes in the surroundings 
occur when the size of the system and the period of
the bulk phase are incommensurate. 

\begin{figure}[th]
\centering
\includegraphics[scale=1]{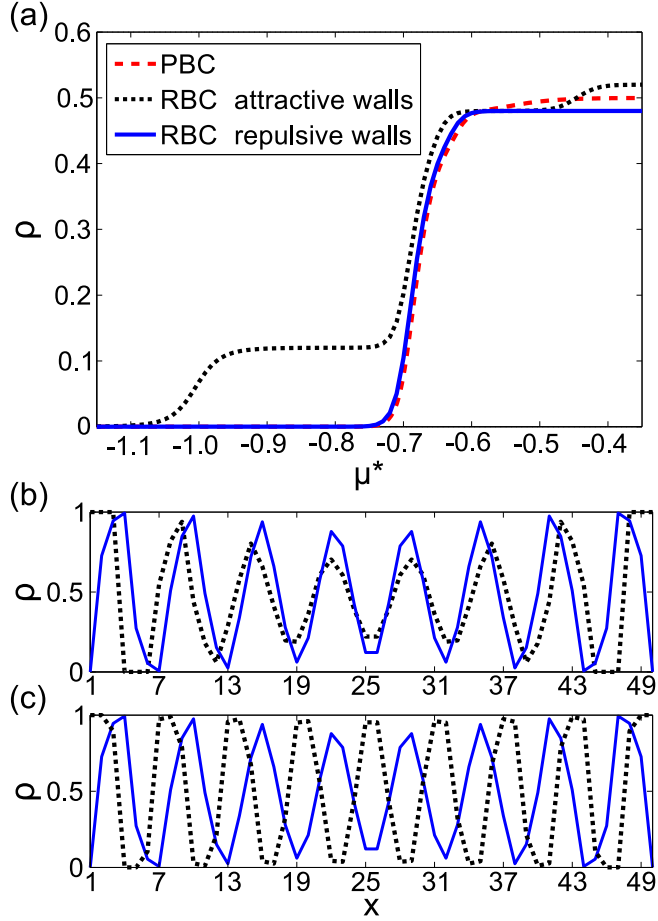}
\caption{Panel (a)  $\rho^*(\mu^*)$ for $J^*=3$, $T^* = 0.03$  and $L=50$  for systems with PBC (dashed line),
RBC with attractive walls (dotted line) and RBC with repulsive walls (solid line).  
The rapid change of density  at $\mu^* \approx -1$ corresponds
to the adsorption of particles on the attractive walls.
%In the case of attractive surfces the number of clusters increases by 1 for $\mu^* \approx -0.45 $.
Panels (b) and (c) show density profiles for 
$\mu^* = -0.55$ and $\mu^* = -0.4$ respectively for  attractive (dashed lines) and repulsive (solid lines) surfaces.
Note the change of the number of clusters for $\mu^* \approx -0.45 $ when the walls are attractive. }
\label{inny_okres2}
\end{figure}

Let us focus on the role of the interaction with the confining surfaces. 
The attractive and repulsive surfaces are compared in Fig. \ref{inny_okres2} for a large slit. 
As expected, when the walls are repulsive we do not observe the step in $\rho(\mu^*)$ at $\mu^*\approx -1$ associated with 
the adsorption. 
In the case of short-range interactions with the walls,
the density profiles in the slits with attractive and repulsive surfaces are very similar  
for $-0.75<\mu^*<-0.5$ (Fig. \ref{inny_okres2}b).
This  rather surprising property follows from the fact that 
 even though  the  clusters  do not touch  the repulsive  surfaces, they are located
very close to them. A significant difference between the attractive and repulsive surfaces appears only for $\mu^*>-0.45$ - there
is one more cluster, and one more step in $\rho(\mu^*)$ in the slit  with the 
attractive surfaces for $L = 20,26,32,\ldots$.

\section{Effective interaction between the walls and deformations of elastic containers}

In this section we discuss the effective potential between the confining surfaces separated by the distance $L$.
We first consider walls separated by a fixed distance. Next we assume that the walls are elastic, 
and the change of the wall separation is possible at the cost of elastic energy. When the equilibrium width of
the empty slit, $L_0$,  and the period of the ordered phase 
do not match, the elastic energy and the fluid-induced stress are in competition. We ask how 
the equilibrium width of the slit filled with the inhomogeneous fluid differs from $L_0$.

\subsection{The case of fixed distance between the confining walls}
\begin{figure}[th]
\centering
\includegraphics[scale=1]{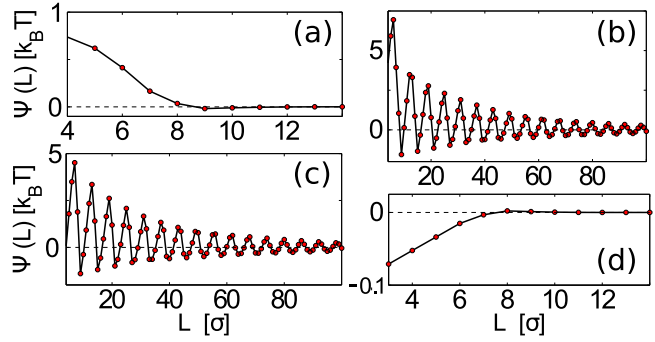}
\caption{$\Psi(L)$ for $J^*=3$ and $T^* = 0.2$ for different values of the chemical potential $\mu^*$ and both walls attractive.
A) $\mu^* = -1$  B) $\mu^* = 0$,  C) $\mu^* = 4$,  D) $\mu^* = 5$. $L$ is in units of the particle diameter $\sigma$}
\label{Psi_ow}
\end{figure}

%for the chemical potential corresponding to the vacuum, the dense phase and the periodic distribution of clusters at $T^*=0$.
%We are interested in the dependence of the decay rate, the amplitude, the wavelength and the phase of $\Psi(L)$ on the chemical potential. 
The exact results for the effective potential between the confining walls separated by a fixed distance, $\Psi(L)$, are presented in Fig. \ref{Psi_ow} for the chemical potential 
corresponding to the GS stability of the vacuum, the periodic phase and the dense phase (compare Fig.\ref{gspbc}).
 Note that the confined fluid leads to repulsion or attraction between the walls  when the dilute
 or the dense pseudo-phase is stable in the bulk respectively. The repulsion may follow from the adsorption
 of the clusters at the surfaces, 
 since the clusters repel each other. The oscillations of $\Psi(L)$ are present if the periodic distribution of clusters
 is thermodynamically preferred. 
These oscillations should be interpreted as follows: the minima of $\Psi(L)$ correspond to the system sizes commensurate
with the periodic structure, therefore if we would allow the system to shrink or expand, then in order
to suppress the internal stress the system would change its size to the value corresponding to the nearest minimum 
of $\Psi(L)$. The bigger is the slope of the oscillations, the stronger is the effective force leading to the nearest 
minimum of $\Psi(L)$. 

\begin{figure}[th]
\includegraphics[scale=1]{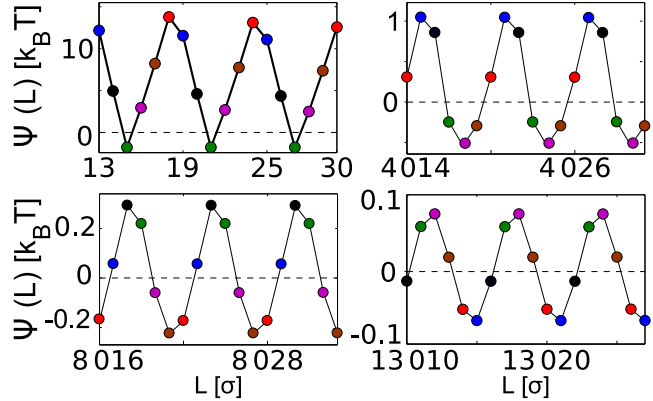}
\caption{$\Psi(L)$  for $J^*=3$, $T^* = 0.2$ and  $\mu^* = 2$ for non-interacting walls. 
$L$ is in units of the particle diameter $\sigma$}
\label{Psi_long}
\end{figure}

For large $L$ the decay rate of $\Psi(L)$, $\xi$, is equal to the bulk correlation length (see Eq.(\ref{psi_ap})).
In Ref. \cite{pekalski:13:0} it was shown  that the correlation length in the considered model 
can be a few orders of magnitude larger than the molecular size for $\mu^*$ corresponding to the 
 stability region of the periodic phase on the GS ($-2/3<\mu^*<14/3$ for $J^*=3$).
%The  potential is of physical significance when it is comparable with $k_BT$, and the amplitude also plays an important role. 
In Fig. \ref{Psi_long} we show that  
for $\mu^*=2$, where $\xi$ takes the maximum, $\Psi(L)\sim 0.1 k_BT$ 
%is of significant strength 
even for system sizes 4 orders of magnitude larger than the particle diameter.

\subsection{The case of elastic confining walls}
\begin{figure}[h]
    \begin{center}
      \includegraphics[scale=1.0]{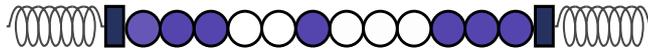}
      \caption{Illustration of the system with elastic walls with spring constat $k$. 
      }
      \label{spring}
    \end{center}
  \end{figure}

\begin{figure}[h]
    \begin{center}
      \includegraphics[scale=1]{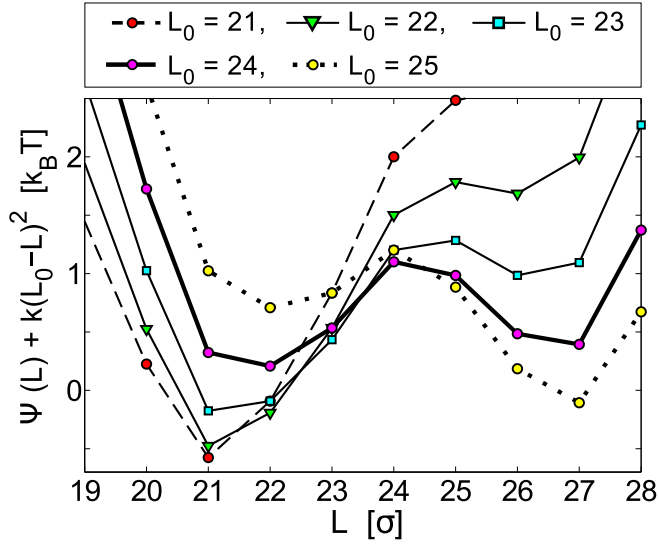}
      \caption{The sum of the elastic energy of the confining boundaries and the effective interaction induced
      by the confined self-assembling system
       for different equilibrium width of the empty slit $L_0$. $J^*=3$, $T^* = 0.5$, $\mu^* = 2$,
      $h^*_1=h^*_L=-1$, and the spring constant $k = 0.1 k_BT/\sigma^2$, where $\sigma$ is the particle diameter. }
      \label{psi_spring}
    \end{center}
  \end{figure}

We assume that the width $L$ of the slit can oscillate around  $L=L_0$, where $L_0$ is the equilibrium width 
in the absence of particles
inside the pore. This 
oscillation can be controlled be a harmonic potential energy  $U_w(L) = k \cdot (L - L_0)^2$ (see Fig. \ref{spring}).
Next we assume that when the slit is in contact with the reservoir of particles, and the chemical potential $\mu^*$ and
 temperature $T^*$ are fixed, then in mechanical equilibrium the sum of $U_w(L)$ and the particle-induced effective potential
 $\Psi(L)$ takes the minimum. We should note that similar assumptions lead to correct prediction of swelling of microporous carbons 
 induced by adsorption of argon~\cite{kowalczyk:08:0}. Here we make a similar assumption for larger particles and 
 system sizes, and softer confining surfaces. 
 %In particular, the model may describe globular charged proteins between lipid bilayers separated by a distance $L\gg 1$.
In Fig. \ref{psi_spring} we present the sum of $\Psi$ and $U_w$ as a function of the system size. Note that when $L_0$ corresponds
to the maximum of $\Psi(L)$, i.e. to a large stress induced by the confined fluid, 
then $U_w(L)+\Psi(L)$ may have two minima of very similar depth for wall separations that differ approximately by the period of the bulk structure. 
The number of clusters in these two states differs by one. As can be seen in Fig.\ref{psi_spring}, the barrier between the two minima
is of order of $k_BT$ for the assumed elastic constant $k=0.1 k_BT/\sigma^2$. 
%Thus, in our bistable system the number of clusters and the width of the slit can change because of thermal fluctuations.

  \begin{figure}[h]
    \begin{center}
      \includegraphics[scale=1]{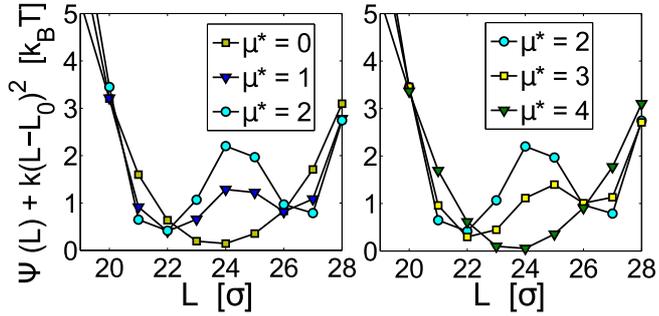}
      \caption{The sum of the elastic energy of the confining boundaries and the effective interaction induced
      by the confined self-assembling system for various values of $\mu^*$. $J^*=3$, $T^* = 0.5$, $h^*_1=h^*_L=-1$,  $L_0=24$, 
      and the spring constant $k = 0.1 k_BT/\sigma^2$.
     }
      \label{psi_barrier}
    \end{center}
  \end{figure}

  In Fig.\ref{psi_barrier} we show how the bistability appears when the chemical potential changes from $\mu^*=0$ or $\mu^*=4$
  towards $\mu^*=2$. The barrier between the two minima decreases for increasing $|\mu^*-2|$. Thus, by changing the concentration of particles 
  in the surroundings we can change the hight of the barrier and induce or suppress the jumps between the two widths of the confined system.

\section{summary}

We have solved exactly the 1d model of a system interacting with the SALR potential in  slits of various widths. The distribution
of the particles in confinement and the effective potential between the confining surfaces have been calculated 
for different values of the chemical potential, from dilute to dense systems in the bulk. We paid particular attention
to $\mu^*$ corresponding to inhomogeneous distribution of the particles in the bulk.
We also obtained $\rho(\mu^*)$ for 
various system sizes and different short-range interactions with the surfaces. We paid particular attention 
to the system sizes incommensurate 
with the typical distance between the clusters or layers in the bulk. 

The most interesting result is the bistability of the system confined by elastic walls (Fig.\ref{psi_spring}). 
The bistability occurs when the 
width in the absence of particles corresponds to the largest stress in the confined self-assembling system. The system choses 
$n$ or $n+1$ layers with almost equal probability. The size difference between
the two cases is similar to the period of the bulk structure that in the case of colloids 
can be as large as hundreads of nanometers or even micrometers. 
%,and is much larger than the size of the particles. 
Similar phenomenon occurs in very narrow slits when the width is such that $n$ and $n+1$ 
atomic layers of the adsorbed gas are equally probable. The size difference, however, is of order of an angstrom. 
An interesting property is the possibility of inducing or suppressing the bistability by changing the chemical
potential, i.e. the concentration of particles in the surroundings. 

The confined self-assembling system behaves as a soft  elastic material itself (Fig.\ref{Psi_ow}), and the bistability takes place when 
its elastic constant is similar to the elastic constant of the boundaries. Such soft boundaries are formed in particular 
by biological membranes.

Another interesting result is the dependence of the deformations in the confined system on the conditions in the surroundings. 
We found that by changing $\mu^*$ we induce changes in the number and size of the layers in the confined system.
These structural changes are reflected in
``steps'' in $\rho(\mu)$. In order to help to interpret simulation results we obtained exact expressions for $\rho(\mu^*)$ 
in the case of PBC and various system sizes. We obtained steps in $\rho(\mu^*)$ corresponding to the change of the type of defects
resulting from the incommensurability.
%of the system size and the period of the bulk structure. 
In order to avoid misinterpretation 
of these steps
as phase transitions, one should verify if the steps disappear for some (commensurate) system sizes, 
and if their heights decays as $\sim 1/L$ 
for increasing system size $L$. 

Recently close similarity between the bulk properties of the SALR and the amphiphilic systems has been
demonstrated in Ref.~\cite{ciach:13:0,pekalski:14:1}.
Based on this similarity we may expect that our results concern also  amphiphilic systems in confinement,
but this expectation should be verified.

Finally, we should note that our exact results concern open systems in contact with a particle reservoir. 
Recently hard discs confined by a ring of particles trapped in holographic optical tweezers, which form
a flexible elastic wall 
were investigated \cite{williams:13:0}. For a fixed number of confined particles a  bistable state of a hexagonal
structure and concentrically
layered fluid mimicking the shape of the confinement was found. This phenomenon has some similarity to
our bistability, since in both cases the adaptive confinement plays a crucial role. However,
the fixed number of confined particles may alter the properties of the system confined between adaptive boundaries. 
Some of the lipid bilayers in living cells are permeable for proteins, 
while some other ones are not. In a forthcoming paper we shall compare the open and closed confined systems
with the same average number of particles. 

\clearpage
\section{APPENDIX}
\subsection{Partition function}
Since the range of particle-particle interactions is $3$, we introduce boxes consisting 
of three neighboring lattice sites. For the system of size $L = 3N+j$, where $j = 0,1,2$,
the boxes can be labeled by integer $k = 1,2,\ldots N$.
The microstates in the $k$-th box are  
 \begin{eqnarray}
\label{S(k)}
\hat S(k)=(\hat\rho(3k-2),\hat\rho(3k-1),\hat\rho(3k)). 
 \end{eqnarray}
 For $N>1$ ($L\ge 6$) the Hamiltonian can be written in the form
 \begin{eqnarray}
 \label{Ha}
H^*[\{\hat\rho\}] =  \ro(1)h_1^*+ \ro(L)h_L^* + H^*_j[\hat S(N)] +  \sum_{k=1}^{N-1} H^*_t[\hat S(k),\hat S(k+1)].
 \end{eqnarray}
 where 
 \begin{eqnarray}
 H^*_t[\hat S(k),\hat S(k+1)]=\sum_{x=3k-2}^{3k}\big[
-\hat\rho(x)\hat\rho(x+1)+J^*\hat\rho(x)\hat\rho(x+3)-\mu^*\hat\rho(x)
\big].
\end{eqnarray}
contains the interaction between two neighboring boxes and the chemical potential term in the first box,
\begin{displaymath}
H^*_j[\hat S(N)]\!=\! \left\{ \begin{array}{ll}
\!  - ( \sum_{i = 0}^{1} \ro(3N\!-\!i)\ro(3N\!-\!i-\!1))-\mu^* (\sum_{i = 0}^{2} \ro(3N-i)) & \textrm{if $j\! =\! 0$}\\
  -( \sum_{i = 0}^{2} \ro(3N\!+\!1 - i)\ro(3N\!-\!i))+J^*  \ro(3N\!-\!2)\ro(3N\!+\!1) & \textrm{if $j\! =\! 1$} \\
-\mu^* (\sum_{i = 0}^{3} \ro(3N+1-i))  &\\
  -( \sum_{i = 0}^{3} \ro(3N\!+\!2 - i)\ro(3N\!+1-i))+ & \textrm{if $j \!=\! 2$}\\
J^* (\sum_{i = 0}^{1} \ro(3N\!-\!2+i)\ro(3N\!+\!1+i))-\mu^* (\sum_{i = 0}^{4} \ro(3N+1-i)) &
\end{array} \right.
\end{displaymath}
contains the particle-particle interactions between the particles which occupy the sites within the 
$N$-th box, and in addition the interactions between the particles at the sites labeled $3N\!+\!1$ and $3N\!+\!2$
 (if such sites exist for given $L$).
Finally, $\rho(1)h_1^*$ and $\rho(L)h_L^*$ are
the  energies of interaction between the particles and the two walls. For $N=1$ the Hamiltonian does not
contain the last term in (\ref{Ha}). We consider only $N >1$ in this work. 
In order to find the partition function of the system we introduce a $8\times 8$  transfer matrix ${\bf T}$ with the matrix elements  
\begin{eqnarray}
\label{T}
{\bf T}(\hat S(k),\hat S(k+1))\equiv e^{-\beta^* H^*_t[\hat S(k),\hat S(k+1)]}.
\end{eqnarray}
The partition function in terms of the transfer matrix has the following form
\begin{eqnarray}
\Xi =\! \sum_{\hat S(1)}  \sum_{\hat S(N)}'  e^{\beta^* \ro(1) h^*_{1} } {\bf T}^{N-1}[\hat S(1),\hat S(N)] e^{\beta^* \ro(L) h_L^*}e^{\beta^* H^*_j[\hat S(N)]},
\end{eqnarray}
where $\sum_{\hat S(N)}'$ denotes
\begin{displaymath}
\sum_{\hat S(N)}'= \left\{ \begin{array}{ll}
\sum_{\hat S(N)}  & \textrm{if $j\! =\! 0$}\\
 \sum_{\hat S(N)} \sum_{\ro(3N\!+\!1)}  & \textrm{if $j\! =\! 1$}\\
 \sum_{\hat S(N)} \sum_{\ro(3N\!+\!1)} \sum_{\ro(3N\!+\!2)}  & \textrm{if $j \!=\! 2$}
\end{array} \right.
\end{displaymath}
We transfer ${\bf T}$ to the base in which it is diagonal and the matrix elements of ${\bf T}^{N-1}$ 
can be easily expressed by the sum over the eigenvalues $\lambda_k$ and the matrix elements $P_k(\hat S(n))$ 
of the matrix transforming ${\bf T}$ to its eigenbasis
\begin{eqnarray}
{\bf T}^{N-1}(\hat S(n),\hat S(m))&=&  \sum_{k=1}^8P_k(\hat S(n))\lambda_k^{N-1}P_k^{-1}(\hat S(m)).
\label{lam}
\end{eqnarray}
Hence the partition function is
\begin{eqnarray}
\Xi = \sum_{\hat S(1)} \sum_{\hat S(N)}'    \sum_{k=1}^{8} e^{\beta^* \ro(1) h^*_{1} } P_k(\hat S(1))\lambda_k^{N-1}P_k^{-1}(\hat S(N))e^{\beta^* \ro(L) h_L^*}e^{\beta^* H^*_j[\hat S(N)]}.
\label{ss}
\end{eqnarray}

\subsection{Average density at a given site}
The framework of the transfer matrix allows us to find a formula for average density at the site $x =3n+l$, 
where $n$ is the number of the triple to which the $x$-th site belongs and $l = 1,2,3$ is 
the label of the site within the triple. For $1<n<N$ the average density at the $x$-th site is
\begin{eqnarray}
\langle\ro (x) \rangle = \frac{1}{\Xi}  \sum_{\hat S(n)} \sum_{\hat S(1)}  \sum_{\hat S(N)}'
e^{\beta^* \ro(1) h^*_{1} } {\bf T}^{n}(\hat  S(1),\hat  S(n)) \ro(x)  {\bf T}^{N-(n+1)}(\hat  S(n), \hat S(N))
e^{\beta^* \ro(L) h^*_L}e^{\beta^* H^*_j[\hat S(N)\!]}.
\end{eqnarray}
In terms of the eigenvalues it takes the form
\begin{eqnarray}
\langle\ro(x)\rangle =  \frac{1}{\Xi}  \sum_{\hat S(n)}  \ro(x) \!\left( \sum_{\hat S(1)}   e^{\beta^* \ro(1) h^*_{1} }
\sum_{k=1}^{8} P_k(\hat S(1)) \lambda_k^{n} P_k^{-1}(\hat S(n))  \right)\cdot  \nonumber \\ 
\left( \sum_{\hat S(N)}' e^{\beta^* \ro(L)h^*_L} e^{\beta^* H^*_j(\hat S(N)\!)}\sum_{k=1}^{8}  P_k(\hat S(n))
\lambda_k^{N-(n+1)} P_k^{-1}(\hat S(N))  \right).
\label{avdenex}
\end{eqnarray}
%In order to obtain an asymptotic expression for $N\to\infty$ for the average density  $\langle\ro(x)\rangle$ 
 We introduce the notation:
\begin{eqnarray}
a_k + i b_k& \equiv&\sum_{\hat S(1)}   e^{\beta^* \ro(1) h^*_{1} }  P_k(\hat S(1)) P_k^{-1} (\hat S(n)),\\
c_k + i d_k &\equiv&  \sum_{\hat S(N)}' e^{\beta^* \ro(L)h^*_L} e^{\beta H^*_j(\hat S(N)\!)}P_k(\hat S(n)) P_k^{-1} (\hat S(N))
% Z_k e^{i z_k} &\equiv& \frac{\lambda_k}{\lambda_1},\\
\end{eqnarray}
where $\lambda_1 \in \mathbb{R}$ is the eigenvalue with the largest absolute value and $i = \sqrt{-1}$.
The dependence of $a_k, b_k, c_k $ and $d_k$ on $\hat S(n)$ is not indicated for clarity of notation. 
The parameters $c_k $ and $d_k$ depend also on $j=L\mod 3$.
Then eq. (\ref{avdenex}) takes form
\begin{eqnarray}
\langle\ro (3n+l)\rangle =  \frac{\lambda_1^{N-1}}{\Xi}  \sum_{\hat S(n)}  \ro (3n+l)
\!\left( \sum_{k=1}^{8} \left(\frac{\lambda_k}{\lambda_1}\right)^n  (a_k + i b_k)  \right) 
\left( \sum_{k=1}^{8}  \left(\frac{\lambda_k}{\lambda_1}\right)^{N-n-1} (c_k + i d_k)  \right).
\end{eqnarray}

Our aim is to obtain an asymptotic expression for $\langle\ro (x)\rangle$ for $N \to \infty$ and $n \sim N/2$.
We sort the eigenvalues  in the descending order of their absolute values and
 neglect in eq.(\ref{avdenex}) all the eigenvalues except from  the first 3  of them.
 %The asymptotic expression can be easily obtained when the 4 largest eigenvalues are well separated. 
 We limit ourselves to the two cases: 
 1) $\lambda_2 = \bar \lambda_3 = |\lambda_2| e^{i\lambda}$  and 2) $\lambda_2, \lambda_3 \in \mathbf{R}$ 
 with  $|\lambda_3 / \lambda_2|^n\ll 1$ for $n\gg 1$. 

If $\lambda_2 = \bar \lambda_3$ then after some algebra we obtain
\begin{eqnarray}
\label{si}
\langle\ro (3n+l)\rangle
&\simeq&\frac{\lambda_1^{N-1}}{\Xi}  \sum_{\hat S(n)}  \ro(3n+l) 
\Big( \! a_1 c_1 +  2 c_1\left(\frac{|\lambda_2|}{\lambda_1}\right)^{n} 
(a_2 \cos(n \lambda) -  b_2 \sin (n \lambda)) 
+  \\
\nonumber
&&  2 a_1  \left(\frac{|\lambda_2|}{\lambda_1}\right)^{N-n-1} 
 ( c_2\cos((N\!\!-\!\!n\!\!-\!\!1)\lambda) - d_2 \sin((N\!\!-\!\!n\!\!-\!\!1)\lambda)) \Big)
\end{eqnarray}
In deriving (\ref{si}) we took into account that 
$\left(\frac{|\lambda_2|}{\lambda_1}\right)^n \cdot \left( \frac{|\lambda_2|}{\lambda_1}\right)^{(N\!-\!n\!-\!1)} 
\ll \left(\frac{|\lambda_2|}{\lambda_1}\right)^{N-n-1}$ for $N \gg 1$ and $n \sim N/2$.
Eq.(\ref{si}) can be written in the form (\ref{avdenapp}) with 
$\xi$ defined in Eq.(\ref{xi}), $\lambda$ defined below Eq.(\ref{xi}), and with the following expressions for the
remaining parameters:
\begin{eqnarray}
\bar{\rho} &\equiv& \frac{\lambda_1^{N-1}}{\Xi} \sum_{\hat S(n)}  \ro(3n+l)  a_1c_1, 
\end{eqnarray}
 \begin{displaymath}
A_1(l) = \left\{ \begin{array}{ll}
 w_2 & \textrm{if $\lambda_2, \lambda_3 \in \mathbf{R}$ and  $|\lambda_3 / \lambda_2|^n\ll 1$}\\
\frac{w_2}{\cos \theta_1 (l)}  & \textrm{if $\lambda_2  = \bar\lambda_3 $}
\end{array} \right.
\end{displaymath}

\begin{displaymath}
A_L(l) = \left\{ \begin{array}{ll}
w_4  & \textrm{if $\lambda_2, \lambda_3 \in \mathbf{R}$ and  $|\lambda_3 / \lambda_2|^n\ll 1$}
\\
\frac{w_4 \exp{(3/\xi)}}{\cos \theta_L (l)}  & \textrm{if $\lambda_2  = \bar\lambda_3 $}
\end{array} \right.
\end{displaymath}
\begin{displaymath}
 \theta_{1}(l) \equiv \arctan \frac{w_3}{w_2}, \hskip1cm \theta_{L}(l)\equiv \arctan \frac{w_5}{w_4} - \lambda
\end{displaymath}
where
\begin{eqnarray}
{w}_2 \equiv \frac{2\lambda_1^{N-1}}{\Xi}  \sum_{\hat S(n)}  \ro(3n+l)  a_2 c_1,
{w}_3& \equiv& \frac{\lambda_1^{N-1}}{\Xi}  \sum_{\hat S(n)}  \ro(3n+l)  b_2 c_1, \\
\quad
{w}_4 \equiv \frac{2\lambda_1^{N-1}}{\Xi} \sum_{\hat S(n)}  \ro(3n+l)  a_1 c_2,
{w}_5& \equiv& \frac{2\lambda_1^{N-1}}{\Xi}  \sum_{\hat S(n)}  \ro(3n+l)  a_1 d_2, \quad
\end{eqnarray}

The above asymptotic expressions are not valid  when $\lambda_2$ and $\lambda_3$ are both real, 
and $|\lambda_3 / \lambda_2|^n =O( 1)$.
  For the range of parameters studied in this article, however, $\lambda_2$ and $\lambda_3$ are complex conjugate numbers.

\subsection{Surface tension and effective interaction between the confining walls}

The grand thermodynamic potential for the bulk system  of the size $L = 3N +j$, where $j = 0,1,2$, and $N\to\infty$ is
\begin{equation}
\label{Omb}
\beta\Omega_{bulk} \simeq_{N\to\infty} - \frac{L}{3N} \ln \lambda_1 ^{N} = -  \ln \lambda_1 ^{N} - \frac{j}{3} \ln \lambda_1 .
\end{equation}

The impact of the system geometry and the particle-wall interactions can be expressed by an excess grand potential 
$\Omega_{ex} \equiv \Omega - \Omega_{bulk}$.
We obtain the grand potential $\Omega$ of the confined system using eq. (\ref{ss}):
\begin{equation}
\beta\Omega = -  \ln \Xi = -  \ln \Big( \sum_{k=1}^{8} \lambda_k^{N-1} C_k(j)  \Big),
\end{equation}
where
\begin{equation}
\label{Ckj}
 C_k(j)=  \sum_{S(1)} \sum_{\hat S(N)}'   e^{\beta^* \ro(1) h^*_{1} }  P_k(\hat S(1))  P_k^{-1}(\hat S(N)) e^{\beta^* \ro(L) h^*_{L} } e^{\beta^* H^*_j(\hat S(N))}.
\end{equation}
From (\ref{Omb})-(\ref {Ckj}) we obtain
\begin{eqnarray}
\label{Oo}
\beta\Omega_{ex} &\simeq&
%- \frac{1}{\beta} \ln \Big( \sum_{k=1}^{8} \lambda_k^{N-1} C_k(j)  \Big) +\frac{L}{3N\beta} \ln \lambda_1^{N} \\
%&=& - \frac{1}{\beta} \ln \Big( \sum_{k=1}^{8} \lambda_k^{N-1} C_k(j)  \Big) +\frac{1}{\beta} \ln \lambda_1^{N} +\frac{j}{3\beta} \ln \lambda_1\\
%&=& - \frac{1}{\beta} \ln \Big( \sum_{k=1}^{8} \lambda_k^{N-1} C_k(j)  \Big) +\frac{1}{\beta} \ln \lambda_1^{N-1} +\frac{3+j}{3\beta} \ln \lambda_1\\
%&=& -\frac{1}{\beta} \ln \Big( \sum_{k=1}^{8} C_k \frac{1}{\lambda_1} \Big(\frac{\lambda_k}{\lambda_1}\Big)^{N-1} \Big) \\
%&=&\frac{3+j}{3\beta}\ln \lambda_1 - \frac{1}{\beta}  \ln \Big( \sum_{k=1}^{8} C_k(j) \Big(\frac{\lambda_k}{\lambda_1}\Big)^{N-1}  \Big)\\
%&=&\frac{3+j}{3\beta} \ln \lambda_1  -\frac{1}{\beta}  \ln \Big( C_1(j) + \sum_{k=2}^{8} C_k(j)\Big(\frac{\lambda_k}{\lambda_1}\Big)^{N-1}  \Big) \\
%&=&\frac{3+j}{3\beta} \ln \lambda_1  -\frac{1}{\beta}  \ln \Big( C_1(j) \Big( 1 + \sum_{k=2}^{8} \frac{C_k(j)}{C_1(j)}\Big(\frac{\lambda_k}{\lambda_1}\Big)^{N-1}\Big)  \Big) \\
%&=&
\frac{3+j}{3} \ln \lambda_1  -  \ln  C_1(j)  -  \ln \Big( 1 + \sum_{k=2}^{8} \frac{C_k(j)}{C_1(j)}\Big( \frac{\lambda_k}{\lambda_1}\Big)^{N-1}\Big)  
%&=&\frac{1}{\beta} \ln \lambda_1 + \frac{j}{3\beta} \ln \lambda_1  -\frac{1}{\beta}  \ln  C_1(j)  -\frac{1}{\beta}  \sum_{k=2}^{8} \frac{C_k(j)}{C_1(j)}\Big( \frac{\lambda_k}{\lambda_1}\Big)^{N-1} \\
%&+&  \frac{1}{\beta}  \ln  C_1(0)  -\frac{1}{\beta}  \ln  C_1(0) \\
%&=& \Big[ \frac{1}{\beta} \ln \lambda_1 -\frac{1}{\beta}  \ln  C_1(0) \Big]+ \Big[\frac{j}{3\beta} \ln \lambda_1  -\frac{1}{\beta}  \ln  %\frac{C_1(j)}{C_1(0)}  -\frac{1}{\beta}  \sum_{k=2}^{8} \frac{C_k(j)}{C_1(j)}\Big( \frac{\lambda_k}{\lambda_1}\Big)^{N-1} \Big] 
\end{eqnarray}
The sum of the  surface tensions and the effective potential between the confining surfaces in Eq.(\ref{Omex}) are given by 
\begin{eqnarray}
\label{st}
\beta(\gamma_1 + \gamma_2) &=&  \ln \lambda_1 - \ln  C_1(0) 
\end{eqnarray}
and
\begin{eqnarray}
\beta\Psi(L)= -\ln \Big( 1 + \sum_{k=2}^{8} \frac{C_k(j)}{C_1(j)}\Big( \frac{\lambda_k}{\lambda_1}\Big)^{N-1} \Big)
\label{Psiex}
\end{eqnarray}
\newline
respectively, since we have verified that the sum of the first two terms in Eq. (\ref{Oo})  does not depend on $j$.
In the asymptotic region of $N \to \infty$  the above expression for $\Psi(L)$ takes the asymptotic form given in  Eq.(\ref{psi_ap})
with  $\phi(j) = \phi_2(j) - \lambda$
and with
\begin{displaymath}
A(j) = \left\{ \begin{array}{ll}
C_2^a(j) e^{3/\xi} & \textrm{if $\lambda_2, \lambda_3 \in \mathbf{R}$ and  $|\lambda_3/ \lambda_2|^N\ll 1$}\\
2C_2^a(j) e^{3/\xi} & \textrm{if $\lambda_2  = \bar\lambda_3 $}
\end{array} \right.,
\end{displaymath}
 where   $C^a_k(j) e^{i\phi_k(j)}=  - \frac{C_k(j)}{C_1(j)}$.\newline

{\bf Acknowledgment\newline}
AC acknowledges the financial support by the National Science Center grant 2012/05/B/ST3/03302.
JP acknowledges the financial support by the National Science Center under Contract Decision No. DEC-2013/09/N/ST3/02551.
N.G.A acknowledges financial support from the Direcci\'on General de Investigaci\'on
Cient\'{\i}fica y T\' ecnica under Grants No. FIS2010-15502 and FIS2013-47350-C5-4-R.
JP received funding for the preparation of the doctoral dissertation from the National Science Center in the funding of PhD scholarships
on the basis of the decision number DEC-2014/12/T/ST3/00647.

 %\bibliography{bibliography_10_14}

\end{document}